\documentclass[12pt]{article}
\usepackage[sort&compress,square,comma,numbers]{natbib}
\usepackage{amsmath,amssymb,graphicx,slashed}
\usepackage{bm}
\usepackage{feynmp}
\usepackage{subfigure}

\usepackage[
	colorlinks=true,
	citecolor=black,
	linkcolor=black,
	urlcolor=blue,
	hypertexnames=false]{hyperref}

\makeatletter
\def\endfmffile{%
	\fmfcmd{\p@rcent\space the end.^^J%
		end.^^J%
		endinput;}%
	\if@fmfio
	\immediate\closeout\@outfmf
	\fi
	\ifnum\pdfshellescape>\z@
	\immediate\write18{mpost \thefmffile}%
	\fi}
\makeatother

\newcommand{\arXiv}[2]{\href{http://arxiv.org/pdf/#1}{{\tt #2/#1}}}
\newcommand{\arXivold}[1]{\href{http://arxiv.org/pdf/#1}{{\tt #1}}}

\newcommand{\beq}{\begin{eqnarray}}
\newcommand{\eeq}{\end{eqnarray}}

\begin{document}
\begin{center} 
{\huge \bf Detecting Dark Matter \\ \vspace*{0.25cm} 
 with Aharonov-Bohm \vspace*{0.5cm}} 
\end{center}

\begin{center} 

{\bf John Terning} and {\bf Christopher B. Verhaaren} \\
\end{center}
\vskip 8pt
\begin{center} 
{\it Center for Quantum Mathematics and Physics (QMAP)\\Department of Physics, University of California, Davis, CA 95616} 
\end{center}

\vspace*{0.1cm}
\begin{center} 
{\tt 
 \href{mailto:jterning@gmail.com}{jterning@gmail.com}\,
 \href{mailto:cbverhaaren@ucdavis.edu}{cbverhaaren@ucdavis.edu}}

\end{center}

\centerline{\large\bf Abstract}
\begin{quote}
While the evidence for dark matter continues to grow, the nature of dark matter remains a mystery. A dark $U(1)_D$ gauge theory can have a small kinetic mixing with the visible photon which provides  a portal to the dark sector. Magnetic monopoles of the dark $U(1)_D$ can obtain small 
magnetic couplings to our photon through this kinetic mixing. This coupling is only manifest below the mass of the dark photon; at these scales the monopoles are bound together by tubes of dark magnetic flux. These flux tubes can produce phase shifts in Aharonov-Bohm type experiments. We outline how this scenario might be realized, examine the existing constraints, and quantify the experimental sensitivity required to detect magnetic dipole dark matter in this novel way.
\end{quote}


\section{Introduction}
Galactic rotation curves \cite{Rubin}, the cosmic microwave background, and the Bullet Cluster \cite{Clowe:2006eq} all provide compelling evidence for dark matter (DM). Despite this wealth of gravitational information, the particle nature of the DM is unknown. The range of possibilities is vast, so we should exploit all available technologies to probe the dark sector. One experimental technique that has not been used in this endeavor is measuring phase shifts from the Aharonov-Bohm (AB) effect~\cite{Aharonov:1959fk}. For most models of DM, there is no AB phase shift, though some effects related to dark sectors have been investigated~\cite{Vachaspati:2009jx,Alavi:2015dmh,Arias:2016vxn}.

A dark $U(1)_D$ sector is detectable if it interacts with the standard model (SM) through a kinetic mixing \cite{Holdom:1985ag} between the visible and dark $U(1)$ field strengths:
\beq
\epsilon e e_D \,F^{\alpha\beta}F^{\alpha\beta}_D~,
\label{mixing}
\eeq
where $e$ and $e_D$ are the visible and dark gauge couplings, respectively. For example, a dark sector with an $SU(2)_D$ gauge group can be broken to $U(1)_D$ when a scalar field, which is a triplet under $SU(2)_D$, gets a vacuum expectation value (VEV) $v_T$. This leads to magnetic monopoles with masses $M\sim4\pi v_T/e_D$~\cite{'tHooft:1974qc} and charges $4\pi/e_D$. If an $SU(2)_D$ scalar doublet gets a VEV $v_D \ll v_T$, then this electric condensate gives the dark photon a mass, $m_D=e_D v_D$. As in a superconductor, the dark monopoles are confined \cite{'tHooftMandelstam}, being connected by tubes of magnetic flux with tension of order $v_D^2$~\cite{Nielsen:1973cs,Bogomol`nyi}. In more sophisticated models \cite{Carlino:2000ff}, monopoles can arise with multiple flavors; alternatively different flavors of monopoles could simply be fundamental particles. An even simpler possibility is that the dark sector only has light electric charges, but the mixing with our photon is through the CP violating operator 
\beq
\varepsilon \, \epsilon_{\mu\nu\alpha\beta}F^{\mu\nu}F^{\alpha\beta}_D~.
\label{cpmixing}
\eeq
Then an $SL(2,Z)$ transformation in the dark sector \cite{Terning:2018lsv} turns dark electric charges into dark magnetic charges and the CP violating mixing (\ref{cpmixing}) into ordinary kinetic mixing (\ref{mixing}). In any of these cases, a flavor non-singlet monopole-antimonopole pair has no annihilation decay channel, and is stable. In an asymmetric DM model one flavor can have a positive magnetic charge excess while another flavor has a negative magnetic charge excess, and charge neutrality ensures that the Universe ends up with stable flavor non-singlet remnants.

In this work we consider dark monopoles as a significant fraction of the DM \cite{Fischler:2010nk,Sanchez:2011mf,Baek:2013dwa,Khoze:2014woa,Hook:2017vyc}.
After a brief review of Lagrangians involving both electric and magnetic charges, we move on to analyzing monopole-antimonopole bound states. We then discuss a variety of existing constraints on our scenario. Next we see how the bound states can be excited when they pass by the Sun and other stars, and then 
 demonstrate how AB phases shifts can arise directly from the passage of DM through the detector. Finally we present our conclusions.

\section{Monopole Interactions}
The formalism of Zwanziger~\cite{Zwanziger:1970hk} is useful for understanding the effects of electric and magnetic charges. The Lagrangian is
\begin{align}
\mathcal{L}=&-\frac{n^\alpha n^\mu}{8\pi n^2}g^{\beta\nu}\frac{4\pi}{e^2}\left(F^A_{\alpha\beta}F^A_{\mu\nu}+F^B_{\alpha\beta}F^B_{\mu\nu} \right)+\frac{n^\alpha n_\mu}{16\pi n^2}\varepsilon^{\mu\nu\gamma\delta} \frac{4\pi}{e^2}\left( F^B_{\alpha\nu}F^A_{\gamma\delta}-F^A_{\alpha\nu}F^B_{\gamma\delta}\right)\nonumber\\
&-A_\mu J^\mu-\frac{4\pi}{e^2}B_\mu K^\mu,\label{e.ZLag}
\end{align}
where $g_{\alpha\beta}=\text{Diag}(1,-1,-1,-1)$ is the Minkowski metric. Here the vectors $A_\mu$ and $B_\mu$ are gauge potentials with local couplings to the electric, $J^\mu$, and magnetic, $K^\mu$, currents, respectively and we have also used the notation
\beq
F^X_{\mu\nu}=\partial_\mu X_\nu-\partial_\nu X_\mu~.
\eeq
The spacelike vector $n^\mu$ ensures that the physical photon, whose degrees of freedom are encapsulated within $A_\mu$ and $B_\mu$, only has two propagating degrees of freedom on shell. While this fixed vector obscures the Lorentz invariance of the theory, physical quantities are expected to be independent of $n^\mu$ when the Dirac charge quantization condition
\beq
qg=\frac{N}{2},
\eeq
is satisfied~\cite{Terning:2018udc}. Here $q$ is the electric charge in units of $e$, $g$ is the magnetic charge in units of $4\pi/e$, and $N$ is an integer.

So far, there is no experimental evidence for magnetic monopoles that couple to the standard model (SM) photon, so we can set $K^\mu=0$. However, a hidden sector with vanishing dark electric current $J_D^\mu=0$ and nonvanishing magnetic current $K_D^\mu$, can be linked to ours by the kinetic mixing given in Eq.~\eqref{mixing}, where
\beq
F_{\mu\nu}=\frac{n^\alpha}{n^2}\left(n_\mu F^A_{\alpha\nu}-n_\nu F^A_{\alpha\mu}-\varepsilon_{\mu\nu\alpha\beta}n_\gamma F^{B\gamma\beta} \right),
\eeq
and similarly for the dark field strength. To simplify the analysis we choose gauges such that $n^\mu$ is the same for both sectors. The fields with diagonal kinetic terms (denoted as $\overline{A}_\mu$ etc) are given by
\beq
\left( \begin{array}{c}
A_\mu\\
A_{D\mu}
\end{array}\right)=&\left( \begin{array}{cc}
\cos\phi+ \epsilon e e_D\sin\phi &-\sin\phi+ \epsilon e e_D\cos\phi\\
\sin\phi & \cos\phi
\end{array}\right)\left( \begin{array}{c}
\overline{A}_\mu\\
\overline{A}_{D\mu}
\end{array}\right),\label{e.Adefs} \\
\left( \begin{array}{c}
B_\mu\\
B_{D\mu}
\end{array}\right)=&\left( \begin{array}{cc}
\cos\phi & -\sin\phi \\
\sin\phi-\epsilon e e_D\cos\phi & \cos\phi+\epsilon e e_D\sin\phi
\end{array}\right)\left( \begin{array}{c}
\overline{B}_\mu\\
\overline{B}_{D\mu}
\end{array}\right),\label{e.Bdefs}
\eeq
to leading order in $\epsilon$. In this basis the visible currents are
\begin{align}
\left( \begin{array}{c}
e\overline{J}_\mu\\
e_D\overline{J}_{D\mu}
\end{array}\right)=&\left( \begin{array}{cc}
\cos\phi+ \epsilon e e_D\sin\phi &\sin\phi \\
-\sin\phi+ \epsilon e e_D\cos\phi & \cos\phi
\end{array}\right)\left( \begin{array}{c}
eJ_\mu\\
e_DJ_{D\mu}
\end{array}\right),\label{e.Jcurrents} \\
\left( \begin{array}{c}
\overline{K}_\mu/e\\
\overline{K}_{D\mu}/e_D
\end{array}\right)=&\left( \begin{array}{cc}
\cos\phi & \sin\phi-\epsilon e e_D\cos\phi  \\
-\sin\phi& \cos\phi+\epsilon e e_D\sin\phi
\end{array}\right)\left( \begin{array}{c}
K_\mu/e\\
K_{D\mu}/e_D
\end{array}\right),\label{e.Kcurrents}
\end{align} Here the angle $\phi$ corresponds to the freedom to choose which linear combination of fields we call the photon. 

This ambiguity is removed when the dark photon gets a mass from an electric condensate,
\beq
\mathcal{L}_\text{photon mass}=-\frac{m_{D}^2}{2}A_D^\mu A_{D\mu}.
\eeq
Then in order to keep the visible photon massless we must have $\phi=0$. Consequently, the SM particles with a photon coupling $eq$ obtain couplings to the dark photon $e_D q_D$ with $q_D=\epsilon e^2q$ (in terms of the conventional normalization for kinetic mixing, $\varepsilon=e\,e_D\epsilon$, they have a coupling $\varepsilon q e$). In addition, dark monopoles with a dark magnetic coupling $g_D 4\pi/e_D$ couple to the visible photon with strength $g_{eff}=\varepsilon g_D 4\pi  /e_D$. Because this charge violates the Dirac charge quantization condition, the magnetic flux strings connecting the monopoles can give rise to physical AB phase shifts. 

Notice that the apparent violation of the Dirac charge quantization condition in each sector is compensated by an effect in the other. Including both sectors one finds a ``diagonal" charge quantization condition \cite{Terning:2018lsv}. In the limit $m_D\to0$ there is the usual freedom to redefine the fields so that the visible photon has no couplings to hidden sector magnetic charges. Consequently, at energies much larger than $m_D$, where the dark photon mass can be neglected, the effect of mixing vanishes. In short, there is no observable AB phase shift on scales smaller than $m_D^{-1}$.

\section{Bound State Analysis}
When the monopole mass is large, $M\gg m_D$, and the magnetic Coulomb interactions between the monopoles are not too large the low-lying states are non-relativistic. The Hamiltonian is approximated by
\beq
H=\frac{p^2}{2 M}-\frac{4\pi g_D^2}{e_D^2 r}e^{-m_Dr} +C \pi\,v_D^2 |r|~,
\eeq
where $C$ is a dimensionless number of order one \cite{Bogomol`nyi} and $r$ is the separation between the monopoles. For large $r$, the Coulomb term can be neglected and the eigenstates of $H$ are Airy functions. Because $MH$ depends on a single dimensionful parameter, $C \pi M v_D^2$, dimensional analysis determines the typical binding energy:
\beq
E\sim \frac{1}{M} \left(C \pi \, M\,v_D^2 \right)^{2/3} = \frac{C^{2/3}  \pi^{2/3} v_D^{4/3}} {M^{1/3}}=0.2\, \text{eV}\,\left( \frac{1\,\text{keV}}{M/C^2}\right)^{1/3}\left( \frac{v_D}{1\,\text{eV}}\right)^{4/3}~,
\label{Etypical}
\eeq
and typical separation
\begin{align}
L\sim& \frac{1} {\left( C \pi\, M \,v_D^2\right)^{1/3}}=14\, \text{nm}\,\left( \frac{1\,\text{keV}}{C M}\right)^{1/3}\left( \frac{1\,\text{eV}}{v_D}\right)^{2/3}~.\label{e.groundR}
\end{align}
Because the low-lying, non-$s$-wave, magnetic dipoles have short lengths, they quickly align with the magnetic fields of the galaxy, Sun, and Earth. Interactions with ordinary magnetic fields through the dipole moment do not lead to experimental bounds because they are highly suppressed by the smallness of both $\varepsilon$ and $L$~\cite{Sigurdson:2004zp}.

For $\alpha_D\equiv4\pi g_D^2/e_D^2$ not too small, the Coulomb term dominates the ground state. Then, the energy states take the usual Hydrogen-like form with
\beq
E_n\approx -\frac{M}{4n^2}\alpha_D^2, \ \ \ \ L_n\approx \frac{2n^2}{M\alpha_D}.\label{e.CoulombEL}
\eeq
The exponential suppression of the Coulombic (Yukawa) potential becomes important at the length scale $m_D^{-1}$, so when $m_D<\alpha_D M/2$ the above results (\ref{e.CoulombEL}) should be accurate for the first few states. By equating $L_{n}$ to Eq.~\eqref{e.groundR} we find the energy level where the two approximations cross over. One finds that when $M\gg m_D$ the ground state is Coulombic, but if $\alpha_D$ is very small then the excited states are not.

The dark sector's cosmological history can easily agree with the successes of standard cosmology: big bang nucleosynthesis and the cosmic microwave background. For sufficiently small kinetic mixing, the dark photon never equilibrates with the SM plasma~\cite{Nelson:2011sf,Evans:2017kti} and the dark sector's temperature is completely independent of the SM. The dark particles can be produced in the early universe~ \cite{Murayama:2009nj} by the reheating following inflation, and we assume that they efficiently annihilate when the dark sector temperature falls below their mass, leaving only a relic of monopoles of one flavor and oppositely charged monopoles of another. As the dark sector cools the dark photon obtains a mass, and the monopoles confine. As in quirk models \cite{Kang:2008ea}, the dipoles quickly de-excite to their ground state by emitting dark photons, which can themselves make up a some fraction of the DM~\cite{Redondo:2008ec}. 

The finite mass of the dark photon may require the final de-excitation to proceed by emission of visible photons. These final decay lifetimes are typically thousands of years or longer due to the small coupling to the visible photon. Late time decays to such photons are in tension with observation~\cite{McDermott:2017qcg}, however, if the dipoles form at an average spacing below the typical size of the first excited state, then there is no de-excitation nor photon emission. We are considering small enough mixing that the two sectors never thermalize, so it is consistent to also assume that the dark confining transition occurs sufficiently early, and no photon production results~\cite{BlancoPillado:1999cy}.

With keV mass monopoles and sub micrometer string lengths, the dominant long-range interaction between dipoles is a van der Waals potential. We parameterize it as
\beq
V_\text{vdW}\sim C_\text{vdW}\frac{g_{eff}^4L^3}{(4\pi)^2\,v_D^2\,r^6}~,\label{e.vdw}
\eeq
where $C_\text{vdW}$ is a dimensionless order one number \cite{Holstein,Terning:2019hgj} and we assumed that the distance between the bound states, $r$, is much greater than $L$. The four powers of the $\varepsilon$ ensure that scattering is accounted for by the Born approximation, and is negligible for the parameter regions of interest. For $r\gg 1/E$, the interaction falls even faster, as $1/r^7$ \cite{Holstein}, again suppressed by $\varepsilon^4$.

\section{Constraints} 

\begin{figure}[th]
\centering
\includegraphics[width=0.49\textwidth]{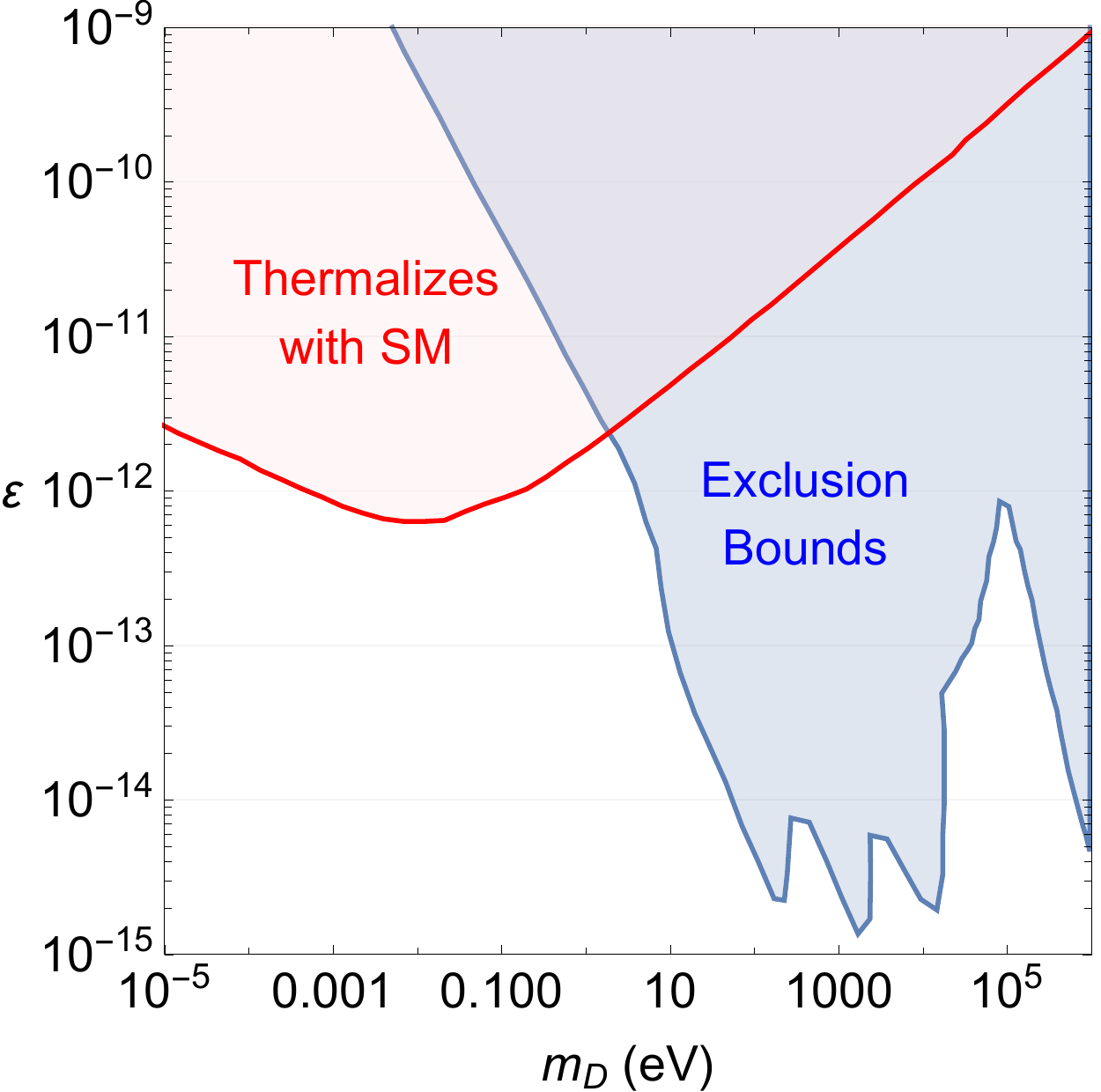}
\includegraphics[width=0.49\textwidth]{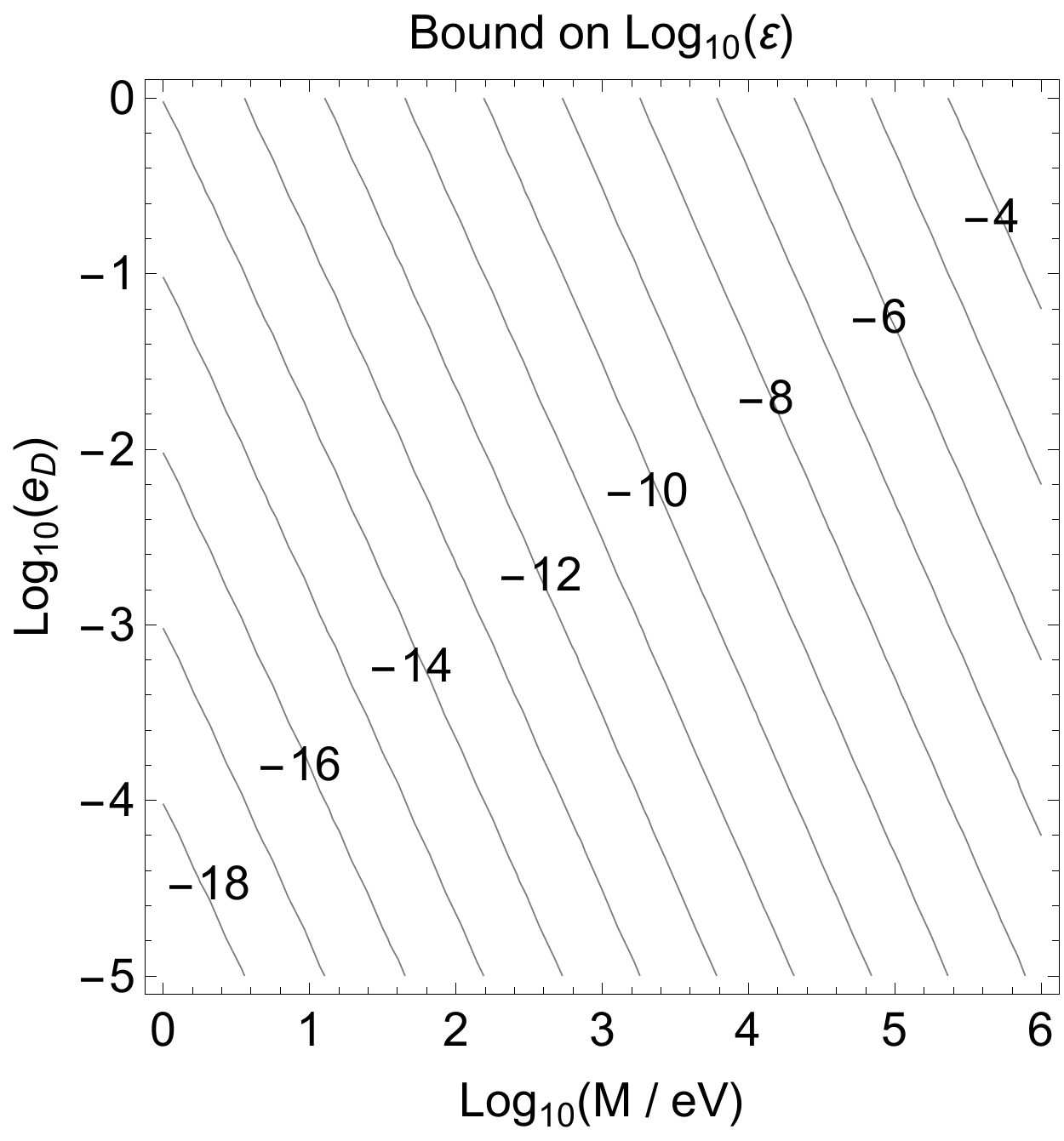}
\caption{\label{f.epsBounds}{\it Left:} Excluded region, in blue, for the kinetic mixing parameter $\varepsilon$ as a function of the dark photon mass $m_D$. These bounds come primarily from stellar cooling and the late decays of dark photons into visible photons. The red region indicates the dark photon thermalizes with the SM~\cite{Nelson:2011sf}. {\it Right:} Bound on $\text{Log}_{10}\,\varepsilon$ from magnetar lifetimes as a function of $M$ and $e_D$, see~\cite{Hook:2017vyc}.}
\end{figure}

Magnetically charged particles are constrained by magnetar lifetimes as a function of their mass, charge, and the mass of the dark photon~\cite{Hook:2017vyc}. The strong magnetic fields of the magnetar can pair-produce the monopoles and deplete their energy. For dark photon masses below 1 eV the bounds are independent of $m_D$, and weaken as the monopole mass $M$ increases, going from $\varepsilon e/e_D\lesssim 10^{-13}$ when $M=1$ eV to $\varepsilon e/e_D\lesssim 10^{-9}$ when $M=1$ keV, as shown in the right panel of Fig. \ref{f.epsBounds}. 

The bounds on the kinetic mixing parameter $\varepsilon$ (through the coupling of the dark photon to SM particles) are quite stringent, the dominant bounds for the region we consider are shown in Fig. \ref{f.epsBounds}. They arise from new sources of stellar cooling~\cite{Gondolo:2008dd}, and late decays of dark photons into visible photons~\cite{Redondo:2008ec,McDermott:2017qcg}.\footnote{The constraints on $\varepsilon$ have been studied over a large span of dark photon masses both from direct experiment and astrophysical observation, see~\cite{Essig:2013lka,Chang:2016ntp,Bauer:2018onh}. 
There are also many proposed new searches for kinetic mixing effects from excitations in condensed matter systems~\cite{Griffin:2018bjn} and from atomic transitions~\cite{Alvarez-Luna:2018jsb}.} These stellar cooling constraints assume a heavy dark Higgs or a small dark coupling, in other cases the bounds become stronger for smaller dark photon masses~\cite{An:2013yua}.

We are interested in the phase shifts from coherent electron beams passing on either side of a magnetic flux tube joining the monopoles. The separation of the beams must be somewhat larger than $m_D^{-1}$ to generate a phase shift. Past AB apparatuses \cite{experiment} have had characteristic scales $R\sim\mu$m which could probe $m_D\gtrsim$ eV. 
Because the string length, Eq.~\eqref{e.groundR}, decreases as $m_D$ increases, reducing detectability, we focus on $m_D\sim$ eV, where $\varepsilon < 10^{-12}$. However, if the beam separation is extended to longer distances then phase shifts from lighter dark photons can also be measured. For instance, if the beams have a mm separation then the dark photon mass can be as low as meV. 

While this class of models has no milli-electric charged particles, which might lead to additional bounds on $\varepsilon$~\cite{Davidson:2000hf,Vogel:2013raa}, the milli-magnetic charges can lead to similar constraints. We note that in some cases supernova shocks can eject milli-electric charged particles from the galactic disk~\cite{Chuzhoy:2008zy,Dunsky:2018mqs}, however these same shocks can only accelerate milli-magnetic dipoles via field gradients leading to much smaller effects. 

Stellar cooling bounds also apply to millimagnetic particles. Within stars the visible photon acquires an effective mass $m_P$ from the plasma. When kinematically allowed, these massive photons can decay to pairs of particles with small electric or magnetic charge, which provides a new cooling mechanism if these decay products can easily escape the star~\cite{Davidson:1991si}. In our scenario, however, when both the visible and dark sectors give an electric mass to the photon, 
 \beq
\mathcal{L}_\text{photon mass}=-\frac{m_{D}^2}{2}A_D^\mu A_{D\mu}-\frac{m_P^2}{2}A^\mu A_\mu.
\eeq
 the mass eigenstates rotate~\cite{Terning:2018lsv}, which in turn affects the couplings. When the mass terms are rewritten using the fields with diagonal kinetic terms we find
\begin{align}
&-\frac12\overline{A}_\mu\overline{A}^\mu\left[ m_P^2\left(\cos^2\phi+2\varepsilon\cos\phi\sin\phi \right)+\sin^2\phi m_D^2\right]\nonumber\\
&-\frac12\overline{A}_{D\mu}\overline{A}^{D\mu}\left[m_P^2\left(\sin^2\phi-2\varepsilon\cos\phi\sin\phi \right)+\cos^2\phi m_D^2 \right]\nonumber\\
&-\overline{A}_{D\mu}\overline{A}^{\mu}\left[ m_P^2\left(-\cos\theta\sin\theta+\varepsilon\cos2\theta \right)+\cos\theta\sin\theta m_D^2\right],
\end{align}
to leading order in $\varepsilon$. Thus, to eliminate the mass mixing we take
 \beq
 \tan2\phi=\frac{2\varepsilon m_P^2}{m_P^2-m_D^2},
 \eeq
where $\phi$ is the angle appearing in Eqs.~\eqref{e.Adefs}\textendash\eqref{e.Kcurrents}. This angle is small as long as $|m_D^2-m_P^2|\gg\varepsilon m_D^2$, and is otherwise order one. Focusing on the small mixing angle case, to leading order in $\varepsilon$ we have $\cos\phi=1$ and 
\beq
\sin\phi=\frac{\varepsilon m_P^2}{m_P^2-m_D^2}~.
\eeq
Consequently, the physical currents from Eqs.~\eqref{e.Jcurrents} and \eqref{e.Kcurrents} are
\begin{align}
e\overline{J}_\mu&=eJ_\mu+\frac{\varepsilon m_P^2}{m_P^2-m_D^2}e_D J_{D\mu}, & e_D\overline{J}_{D\mu}&=e_DJ_{D\mu}-\frac{\varepsilon m_D^2}{m_P^2-m_D^2}e J_{\mu},\\
\frac{1}{e}\overline{K}_\mu&=\frac{1}{e}K_\mu+\frac{\varepsilon m_D^2}{m_P^2-m_D^2}\frac{1}{e_D}K_{D\mu}, & \frac{1}{e_D}\overline{K}_{D\mu}&=\frac{1}{e_D}K_{D\mu}-\frac{\varepsilon m_P^2}{m_P^2-m_D^2}\frac{1}{e}K_{\mu}~.
\end{align}
Thus, the coupling of the visible photon to the dark monopoles becomes
\beq
\varepsilon\to \varepsilon\frac{m_D^2}{m_D^2-m_P^2},
\eeq
 for $|m_D^2-m_P^2|\gg\varepsilon m_D^2$. When the masses are nearly degenerate the mixing angle between the two photon states becomes $\sim\pi/4$, and the millicharge bound $\varepsilon<10^{-14}$ applies for kinematically producible monopole masses. Horizontal branch stars, red giants, and white dwarfs have average plasma masses ranging from a few keV to 20 keV~\cite{Davidson:2000hf}, so the bounds become
\beq
 \varepsilon\frac{m_D^2}{m_D^2-m_P^2}\lesssim 10^{-14}, \;\text{for monopole masses}\; M\lesssim m_P/2.
\eeq
Note that for $m_D\sim$ eV, this reduces the bound to about $\varepsilon\lesssim 10^{-8}$, making this bound weaker than those shown in Fig.  \ref{f.epsBounds}. When the dark photon mass is larger than $m_P$ there is no such reduction and this constraint can dominate. However, if the monopole mass is increased beyond a few tens of keV this decay is kinematically forbidden, obviating the bound. 

The above constraints show that the largest values of $\varepsilon$ (and hence of magnetic couplings and AB phase shifts) require lighter dark photons, with masses $m_D\lesssim$ eV and somewhat heavier monopoles, with masses $M\sim$ keV. If both the masses are increased the bounds relax, but the string tension increases and the separation lengths contract which diminishes the signal. However, this can be compensated, to some degree, by reducing the size of the detector. Still, with a smaller detector and heavier DM mass, the number of dark dipoles that pass through the detector diminishes.

Galaxy clusters~\cite{Harvey:2015hha} show that DM self-interactions satisfy $\sigma /(2M)\lesssim 0.47\,\text{cm}^2/\text{g}=$
$(13\,{\rm GeV}^{-1})^3$. Reasoning by analogy from atomic scattering shows that the self-interaction cross section of the ground states exceeds these values. However, if the dark dipoles make up only 10\% of the DM then the elastic self-interaction is essentially unbounded~\cite{Fan:2013yva}. Consequently, we assume the monopole DM makes up about 10\% of the cosmological DM. 

\section{Photon Excitation}
The confined monopole ground state is in the  $s$-wave. Hence, it has no dipole moment and no AB effect because there is no net magnetic flux in any direction. However, nearby stars, especially the Sun, can excite a fraction of the DM to states with a dipole moment, providing a detectable signal. From first order perturbation theory, the rate of photon absorption from the ground state $|n\ell m\rangle=|100\rangle$ to an excited state with one unit of angular momentum $|210\rangle$ is 
\beq
R_{1\to 2}=\frac{g^2_{eff}}{3\pi}\frac{\omega^3\left|\langle 100|\bm{r}|210\rangle\right|^2}{e^{\omega/T_s}-1},
\eeq
where $\omega$ is the frequency of the photon with energy equal to the difference between the two states and $T_s=0.5$ eV is the surface temperature of the Sun. We approximate $\langle 100|\bm{r}|210\rangle$ by the displacement of the two monopoles given in Eq.~\eqref{e.CoulombEL} and take the frequency $\omega=E$, also given by Eq.~\eqref{e.CoulombEL}, to find
\beq
R_{1\to 2}\sim\frac{9}{16}\varepsilon^2M\alpha_D^5\left[ \exp\left(\frac{3M\alpha_D^2}{16T_s} \right)-1\right]^{-1}.\label{e.absorbRate}
\eeq
We use the Coulombic values for the ground and first excited states, which means we have chosen $\alpha_D$ large enough that at least two states are Coulomb-like. Equation~\eqref{e.absorbRate} gives the DM photon absorption rate at the surface of the Sun, or at a radius $R_S$. Notice that we need $3M\alpha_D^2/16\lesssim$  eV to have an appreciable number of photons of the correct energy to excite the dipole. 
If the energy splitting between the $\ell=1$ and ground states, see Eq.~\eqref{Etypical}, is larger than $m_D$ then the dipole promptly de-excites by emitting a dark photon. For the dipoles to remain excited for detection on laboratory time-scales we need
\beq
m_D>\frac{3}{16}M\alpha_D^2.\label{e.darkDecay}
\eeq

We are interested in the average number of DM particles which absorb a photon before passing through an AB apparatus on the Earth. This is estimated by considering a straight-line DM trajectory from infinity to the Earth (the origin) given by $vt$ where $v$ is the DM velocity and $t$ is time. This path, which makes an angle $\theta$ with the Sun in the sky, brings the DM through the Sun's photon flux. With the Sun on the $y$-axis a distance $R_A$ away, the distance $R$ between the center of the Sun and the DM is given by $R^2=v^2t^2-2vt R_A\cos\theta+R_A^2$.
The rate in Eq.~\eqref{e.absorbRate} can be rescaled to this distance from the Sun by $R_S^2/R^2$.  Then, the total number of photon absorptions over the time of travel from infinity is
\beq
\int_0^\infty dt\,\frac{R_S^2}{R^2}.
\eeq
We then average over the whole sky to obtain\footnote{We neglect the small effect of excising the Sun from the $\theta$ integral.} $\pi^2R_S^2/(4v R_A) $.
When multiplied by the rate in Eq.~\eqref{e.absorbRate}, this yields the average number of DM which are excited to the $\ell=1$ state by a solar photon before arriving at the Earth's surface. The lifetime of the excited state goes like the inverse of the absorption rate as long as Eq.~\eqref{e.darkDecay} is satisfied, and is typically longer than thousands of years. By multiplying the average rate by the DM number density ($n_{DM}=\rho_{DM}/2M$, where $\rho_{DM}=0.4$ GeV/$\text{cm}^3$) and the DM velocity we find the flux of excited DM at the surface of the Earth:
\beq
F_{DM}\sim f_D\frac{\rho_{DM}}{2M}\frac{R_S^2}{R_A}\frac{9\pi^2}{64}\varepsilon^2M\alpha_D^5\left[ \exp\left(\frac{3M\alpha_D^2}{16T_s} \right)-1\right]^{-1},\label{e.exDMflux}
\eeq
where $f_D$ is the fraction of the DM that is dark magnetic monopoles. Thus, the DM signal is directional, tracking with the position of Sun approximately one month earlier.

Of course, we need not depend on stars to excite the dipoles. Resonant photon cavities  placed around the detector can also produce the required signal. Tuning the cavity frequency would allow one to learn about the excitation spectrum of the DM. 

\section{Aharanov-Bohm Detection}
The AB phase shifts are obtained from the vector potential for magnetic dipole system. For infinite strings Jordan found \cite{Jordan} 
\beq
\vec{A}(\vec{x})=\frac{g_{eff}}{4\pi}\int_\text{String}\frac{d\vec{\ell}'\times(\vec{x}-\vec{x}')}{|\vec{x}-\vec{x}'|^3}~,
\eeq
which can be simply restricted to a finite string length $L$. For a string along the negative $z$ axis the vector potential has only one nonzero component (in cylindrical coordinates)
\beq
A_\phi(\rho,z)=\frac{g_{eff}}{4\pi \rho}
\left[\frac{L+z }{\sqrt{(L+z)^2+\rho^2}}-\frac{z}{\sqrt{\rho^2+z^2}}\right]  .\label{e.Aphi}
\eeq

The Aharonov-Bohm phase for a particle of charge $q$ encircling a region of magnetic flux is
\beq
\Phi_\text{AB}=eq\oint d\vec{x}\cdot\vec{A}~.
\eeq
For simplicity, and since the phase is topological \cite{Terning:2018udc}, we can consider the following set-up: one electron making a semicircular path above the string and one making the mirror path below it. Of course, the dark matter is moving at some velocity $v\sim 300$ km/sec past the detector. We compensate for this, by boosting to the rest frame of the monopoles, and taking the electron paths to be moving in the $z$ direction with velocity $v$: $z(t)=z_0+vt$.
We take the half circular path to have length $\pi R$, with $R$ the distance (in the $x$-$y$ plane) from the string to the path. The time it takes to traverse the path is $t_T=\pi R/v_e$ where $v_e$ is the electron velocity in the lab frame. This yields $\phi(t)=\phi_0+tv_e/R$ and allows us to rewrite Eq.~\eqref{e.Aphi} as
\beq
A_\phi(z)=\frac{gb}{4\pi R}\left[\frac{ L+z}{\sqrt{(L+z)^2+R^2}}-\frac{z}{\sqrt{z^2+R^2}} \right].
\eeq
The phase difference between the two paths is just twice the phase from the first path
\begin{align}
\Phi_\text{AB}&=2\pi qg\left(\frac{L+z_0}{\sqrt{(L+z_0)^2+R^2}}-\frac{z_0}{\sqrt{z_0^2+R^2}} \right)+\mathcal{O}\left(\frac{v}{v_e}\right),
\end{align}
where we have expanded to leading order in $v/v_e\ll 1$. This analysis assumes that the string passed perpendicularly through the area enclosed by the electron beams. This can always be accomplished by aligning the detector with the magnetic field lines of the Earth, with which the dipoles align.

\begin{figure}[th]
\centering
\includegraphics[width=0.49\textwidth]{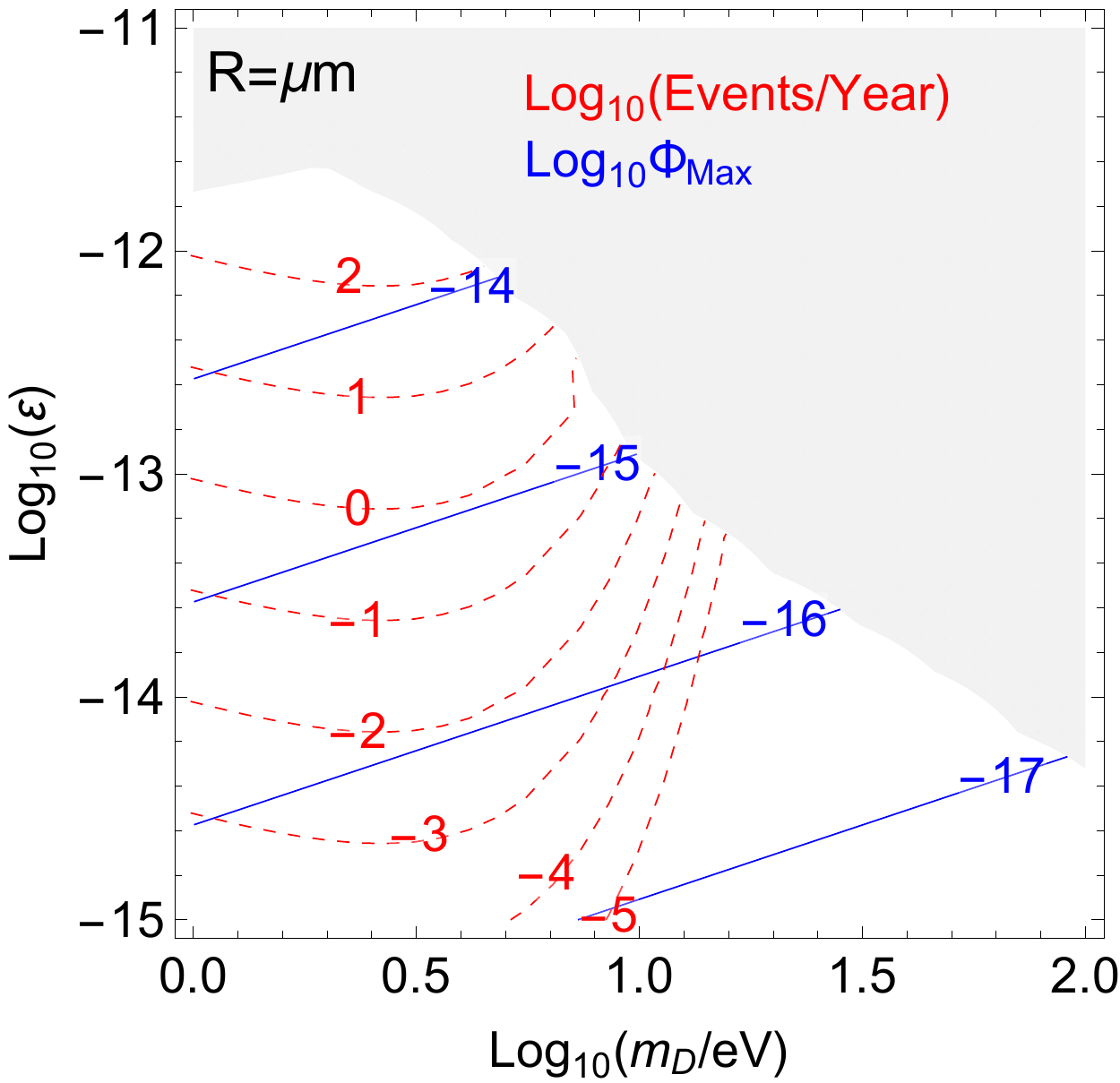}
\includegraphics[width=0.49\textwidth]{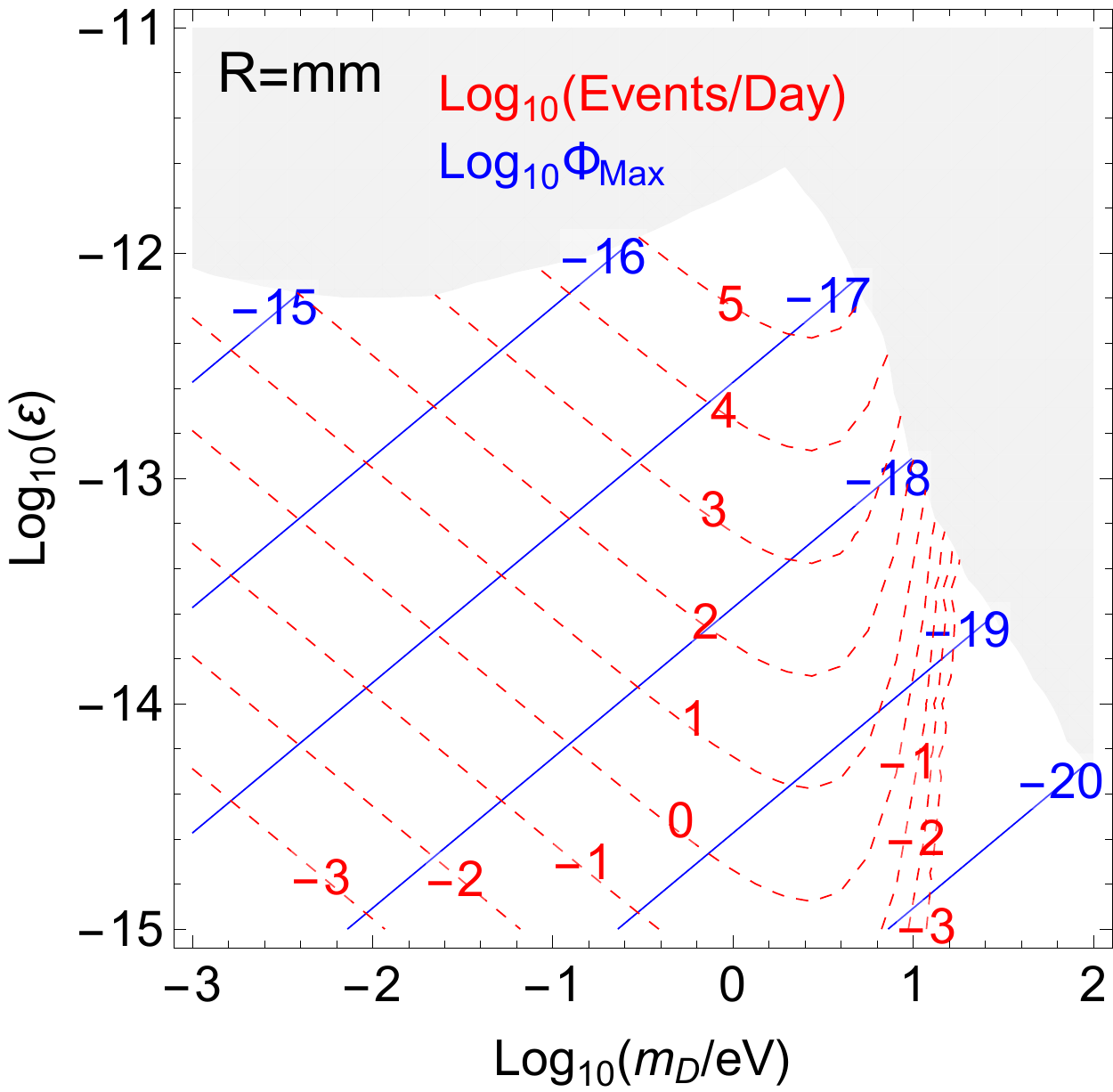}
\caption{\label{f.results} Blue solid contours of maximum AB phase shift and on the left (right) red dashed contours of the expected number of AB event per second (year) from solar excitation (assuming the monopoles make up 10\% of the DM) as a function of the dark photon mass $m_D$ and the kinetic mixing $\varepsilon$. The characteristic size of the detector on the left (right) is $R=1$ mm ($\mu$m) and $\alpha_D=0.07$ (0.005). The monopole mass is $M=$1 keV. The gray region is excluded by dark photon constraints or leads to the dark sector thermalizing withe the SM.}
\end{figure}

Putting all the pieces together, we estimate the maximum phase shift ($z_0=0$) as
\beq
\Phi_\text{max}=2\pi q \frac{e}{e_D}g_D\varepsilon \frac{L}{\sqrt{R^2+L^2}}.
\eeq
With our conventions the charge of the electron is one, and we expect that $g_D$ is also order one. However, the value of $e_D$ is a relatively free parameter, it can be small as long as Eq.~\eqref{e.darkDecay} is satisfied. Typically the characteristic size of the experiment is much larger than the string length. For instance, we need $R\sim \mu$m to enclose the flux tube of a dark photon with eV scale mass. Then, string length estimate in Eq.~\eqref{e.CoulombEL} implies $L\ll R$. We can then estimate the phase shift as
\begin{align}
\Phi_\text{max}\approx& \frac{2qe\varepsilon}{MR}\sqrt{\frac{\pi}{\alpha_D}}\approx 10^{-15} \left(\frac{\varepsilon}{10^{-12}} \right)\left( \frac{10^{-6}\,\text{m}}{R}\right)\left( \frac{1\,\text{keV}}{M}\right)\left( \frac{0.025}{\alpha_D}\right)^{1/2}.
\end{align}
From Eq.~\eqref{e.exDMflux} we find the corresponding flux from solar excitation through a $\mu$m sized detector to be $F_{DM}\sim 3\times10^{-8}\,\text{sec}^{-1}$ yielding one event per year. Figure~\ref{f.results} shows contours of phase shift and expected event flux as a function of $\varepsilon$ and $m_D$ assuming dark dipoles make up 10\% of the DM.

The AB effect was originally verified using modified electron microscopes with $\mu$m resolution with long exposure times \cite{experiment}.  This type of equipment can reach AB phase sensitivities $\sim 10^{-2}$--$10^{-3}$. Modern electron microscopes have demonstrated atomic scale and femtosecond resolutions~\cite{Zewail}, but, as far as we know, have not been employed for AB measurements.
Hopefully this technology can be deployed in the search for dark matter, and the phase sensitivity can be improved to the point where meaningful bounds (or discoveries) are possible.

\section{Conclusion}
We have shown that a sizable fraction of  cosmological dark matter may be composed of the magnetic dipoles coupled to a massive dark photon. A kinetic mixing between the two photons gives these dipoles a small coupling to the visible photon, below the scale of the dark photon mass. We have shown that the constraints on these millimagnetically charged particles come largely from the usual dark photon bounds, though there are new interesting effects to consider.

The small magnetic coupling violates the Dirac charge quantization condition in our sector, allowing the flux tubes joining the monopoles to produce observable phase shifts in AB experiments, constituting a novel DM search strategy. While these shifts are small, the improvements in technology since the last generation of AB experiments make measuring such small effects within the realm of possibility.

\appendix
\section*{Acknowledgments}
We thank Asher Berlin, Hsin-Chia Cheng, David Dunsky, Paul Geiger, Can Kilic, Markus Luty, Tim Tait, and Tien-Tien Yu for helpful discussions. This work was supported in part by the DOE under grant DE-SC-0009999.

\end{document}